\documentclass[prl,twocolumn,superscriptaddress,nofootinbib,longbibliography,tightenlines]{revtex4-2}
\usepackage[svgnames]{xcolor}
\usepackage{amsmath,amssymb,amsthm,bbm}
\usepackage{xr-hyper}
\usepackage{hyperref}
\usepackage{graphicx}
\usepackage{float}
\usepackage{physics}
\usepackage[caption=false]{subfig}
\usepackage[capitalise]{cleveref}
\hypersetup{
  colorlinks,
  citecolor=Salmon,
  linkcolor=Teal,
  urlcolor=Teal}

\newcommand\RKK[1]{{\textsf{\footnotesize{\color{blue}[RKK: #1]}}}}
\newcommand\GM[1]{{\textsf{\footnotesize{\color{red}[GM: #1]}}}}

\begin{document}

\title{Lattice Model For The  Quantum Anomalous Hall Effect in Moir\'e Graphene}
\author{Ahmed Khalifa}
\affiliation{Department of Physics \& Astronomy, University of Kentucky, Lexington, KY 40506, USA}
\author{Ganpathy Murthy}
\affiliation{Department of Physics \& Astronomy, University of Kentucky, Lexington, KY 40506, USA}
\author{Ribhu K. Kaul}
%\affiliation{Department of Physics \& Astronomy, University of Kentucky, Lexington, KY 40506, USA}
\affiliation{Department of Physics, The Pennsylvania State
University, University Park, PA 16802, USA}
\begin{abstract}
Inspired by experiments on magic angle twisted bilayer graphene, we present a lattice mean-field model for the quantum anomalous  Hall effect in a moir\'e setting. Our hopping model thus provides a simple route to a moir\'e Chern insulator in commensurately twisted models. We present a study of our model in the ribbon geometry, in which we demonstrate the presence of thick chiral edge states that have a transverse localization that scales with the moir\'e lattice spacing. We also study the electronic structure of a domain wall between opposite Chern insulators. Our model and results are relevant to experiments that will image or manipulate the moir\'e quantum anomalous Hall edge states. 
\end{abstract}

\maketitle

{\em Introduction:} In a remarkable burst of experimental progress, magic angle twisted bilayer graphene has been found to host a wide range of novel quantum phenomena, from band topology to superconductivity~\cite{Cao2018,Cao20182,doi:10.1126/science.aav1910,doi:10.1126/science.aaw3780,doi:10.1126/science.aay5533,Lu2019,Chen2020}. Notable among the states of matter discovered is the quantum anomalous Hall effect (QAHE), in which time reversal symmetry is spontaneously broken resulting in a quantized Hall conductance~\cite{doi:10.1126/science.aay5533,doi:10.1126/science.aaw3780,Chen2020,Grover2022}. This phenomenon, which will be the focus of this work, brings together in a new setting the two  pillars of the contemporary study of quantum materials: strong correlations and band topology, all in a structure made simply of Carbon. 

The iconic lattice description for the QAHE is Haldane's honeycomb model~\cite{PhysRevLett.61.2015}. Although the model explicitly breaks time reversal symmetry, it may be viewed as a useful mean field model when time reversal symmetry is broken spontaneously by strong interactions. The model starts with spinless electrons hopping on the honeycomb lattice, which results in a linear dispersion at half-filling and at low energies. This is graphene's celebrated realization of the  Dirac equation in which the masslessness is protected by a combination of ${\cal C}_2 {\cal T}$ symmetry where ${\cal C}_2$ denotes rotation by $\pi$ around the $z$ axis and ${\cal T}$ denotes time reversal. Haldane's perturbation consists of an imaginary second neighbor hopping chosen specifically to break ${\cal T}$ but preserves ${\cal C}_2$. Such a perturbation can be shown to guarantee a non-zero Chern number, giving rise to the QAHE. While the $k$-space picture is convenient to establish bulk topology,  the chiral edge states of the Chern insulator are best studied in the real-space ribbon geometry. These edge modes have a transverse localization length on the scale of the spacing $a$ of the underlying honeycomb lattice. Our goal in this work is to construct  a mean field model for the QAHE in moir\'e graphene, formulated on a real-space lattice analogous to Haldane's model. Such a model would  allow  studies of the electronic structure, including the chiral models, of the QAHE state with inhomogeneity in real space, such as that arising from the presence of edges, internal domain walls, electrostatic potentials due to external gates, density gradients, and disorder all of which play an important role in experiments on moir\'e graphene.

\begin{figure}[t!]
\captionsetup[subfigure]{labelformat=empty}
\includegraphics[width=0.48\textwidth]{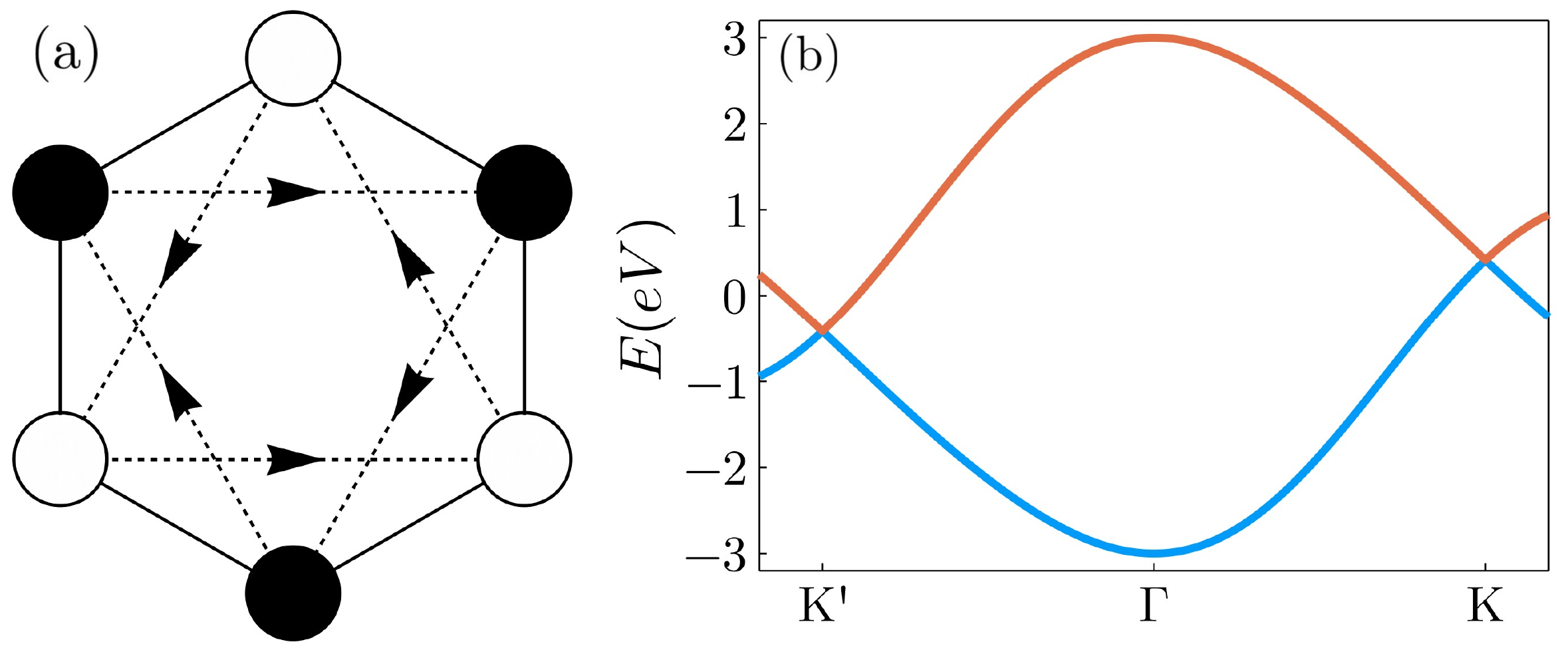}
	%\subfloat[\label{hmh_hopping}]{	
  %\includegraphics[width=0.13\textwidth]{hopping_pattern.pdf}}
  %\hspace{0.05em}
	%\subfloat[\label{hmh_bands}]{%
		%\includegraphics[width=0.23\textwidth]{mh_dispersion.pdf}}
	\caption{\label{fig:hmh} The $\lambda_{\rm mh}$ perturbation of Eq.~(\ref{eq:hmh}) on a single sheet of graphene.  (a) the pattern of imaginary second neighbor hoppings of the $\lambda_{\rm mh}$ term that breaks ${\cal C}_2$ and ${\cal T}$ individually but preserves their combination. This is a modification of the Haldane model in which the two sub-lattices have the imaginary hopping in the same sense; in this ``modified Haldane" model they are in the opposite sense. Since the $\lambda_{\rm mh}$ preserves ${\cal C}_2 T$, it does not act as a mass term. (b) the resulting dispersion of $H_{\rm mh}$ in a section of the graphene Brillouin zone showing that the perturbation preserves the masslessness of the Dirac equation in each valley but raises the energy of one Dirac point with respect to the other. }
\end{figure}

We first review a popular physical picture for the QAHE in moir\'e graphene~\cite{andrei2020graphene,liu2021orbital,PhysRevB.99.075127,PhysRevLett.124.166601,PhysRevLett.124.046403,PhysRevLett.124.187601,PhysRevB.103.075122,PhysRevB.104.115404,PhysRevLett.125.226401,PhysRevLett.124.046403,PhysRevB.103.035427,PhysRevResearch.1.033126,PhysRevResearch.3.013033,PhysRevB.102.035136,PhysRevB.100.085136,PhysRevB.102.045107,PhysRevB.102.035441}. The starting point is the continuum model~\cite{PhysRevLett.99.256802,doi:10.1073/pnas.1108174108} in which moir\'e reconstructions of the band structure at the K and K$^\prime$ valleys are treated independently.  Close to the magic angle the band structure consists of two almost flat bands in each valley that touch at the corners of the moir\'e BZ, a touching protected by ${\cal C}_2 {\cal T}$. Upon applying sub-lattice masses on the two layers (via the alignment with an HBN substrate~\cite{Jung2015,PhysRevB.96.085442,doi:10.1126/science.1237240,PhysRevLett.110.216601,Zibrov2018,Kim2018,PhysRevB.92.155409}), ${\cal C}_2$ is broken, the two bands in each valley are gapped, and carry $\pm 1$ Chern number~\cite{PhysRevB.99.075127,PhysRevLett.124.166601}. By time reversal the opposite valley has exactly the same band structure but with opposite Chern numbers $\mp 1$. The QAHE occurs when the band structure has just enough electrons to fill up one of these four bands, with electron-electron interactions spontaneously breaking time reversal symmetry, resulting in a valley-polarized Chern insulator. In a mean field (Hartree-Fock) picture of the continuum model, we may posit that the energy for the electrons in one valley is raised with respect to the other valley resulting in valley polarization. We note here that it is widely believed that the QAHE state in moir\'e graphene is spin polarized (in addition to the valley polarization just discussed). Since it also does not play a crucial role for the physics we are looking at, in our manuscript  we will henceforth ignore the electron spin, and focus on the orbital route to magnetism.

\iffalse
\begin{figure}[t!]
	\subfloat[\label{hmh_hopping}]{	
  \includegraphics[width=0.13\textwidth]{hopping_pattern.pdf}}
  \hspace{0.05em}
	\subfloat[\label{hmh_bands}]{%
		\includegraphics[width=0.23\textwidth]{mh_dispersion.pdf}}
	\caption{\label{fig:hmh} The $H_{\rm mh}$ perturbation Eq.~\ref{eq:hmh}. The ${\cal C}_2 {\cal T}$ preserving hopping introduced here. The left panel shows the pattern of imaginary second neighbor hoppings that breaks ${\cal C}_2$ and ${\cal T}$ individually but preserves their combination. The right panel shows the resulting dispersion, which demonstrates how the perturbation preserves the masslessness of the Dirac equation in each valley but raises in energy one Dirac touching with respect to the other. }
\end{figure} 
\fi
{\em Lattice Model:} It is not  obvious {\em a priori} how to write a mean field lattice model  that leads to the the QAHE, with the same mechanism as in the continuum mean field picture. Nonetheless, inspired by the continuum picture, we seek a perturbation to the monolayer graphene Hamiltonian that raises the energy in one valley with respect to the other, but maintains the masslessness of the Dirac points in each valley. The perturbation must break both ${\cal T}$ and ${\cal C}_2$ so that the band structures in the two valleys can be different, but must preserve ${\cal C}_2 {\cal T}$ so that the Dirac equation is left massless. Such a term can be constructed by modifying the honeycomb Haldane model, arranging for the sense of the imaginary hopping on the A and B sub-lattices to run oppositely (in the Haldane model they run in the same direction). This is shown pictorially in Fig. (\ref{fig:hmh}),
\begin{equation}
\label{eq:hmh}
H_{\rm mh}= -t\sum_{\langle i,j \rangle}c_{i}^{\dagger}c_{j}+\lambda_{mh}\sum_{\langle\langle i,j \rangle\rangle}\mathrm{i}\nu_{ij}c_{i}^{\dagger}c_{j},    
\end{equation}
where $c_{i}$ denotes the electron's annihilation operator at honeycomb site $i$ and $\nu_{ij}=\pm 1$. The A-A hopping and B-B hopping change sign under both  ${\cal C}_2$ and  ${\cal T}$, but are left invariant under the ${\cal C}_2{\cal T}$. The band structure of monolayer graphene perturbed with $\lambda_{\rm mh}$, shown in the right panel of Fig. (\ref{fig:hmh}), displays the expected behavior, two massless Dirac cones at K and K$^\prime$ but with touching points that are displaced in energy. This model has been introduced previously, and was dubbed the ``modified Haldane model'' in the context of studying anti-chiral edge states in two dimensional fermion systems~\cite{PhysRevLett.120.086603,PhysRevB.87.035427}. While the $\lambda_{mh}$ term does not lead to a Chern insulator in monolayer graphene, we now show that when added to twisted moir\'e graphene it results in the QAHE.

We now incorporate this perturbation into a lattice model of twisted bilayer graphene. In order to work on a moir\'e system on the lattice, we use commensurate twist angles with large moir\'e triangular lattice vectors ${\bf A}_1$ and  ${\bf A}_2$ ($A_M \equiv |{\bf A}_1|=|{\bf A}_2|$) and a smaller moir\'e Brillouin zone~\cite{PhysRevLett.99.256802,PhysRevB.81.161405,PhysRevB.81.165105}. To demonstrate that  $\lambda_{\rm mh}$ indeed results in the QAHE in the moir\'e graphene, we construct a tight-binding model on the commensurate lattice~\cite{T2010,PhysRev.94.1498,PhysRevB.87.205404,PhysRevB.87.075433} for the twisted bilayer graphene system,
\begin{equation}
    \label{eq:h}
    H_{\rm TBG}=\sum_{\mu}H_{\mu,\rm mh}+\sum_{\mu,i}m_{\mu}\sigma_{z}c_{\mu,i}^{\dagger}c_{\mu,i}+\sum_{i,j}t_{ij}^{\perp}c_{t,i}^{\dagger}c_{b,j},
\end{equation}
where $c_{\mu,i}$ is annihilation operator for an electron in layer $\mu=t,b$ and at lattice position $i$. The first ``intralayer" term, defined in Eq. ~(\ref{eq:hmh}), consists of the usual graphene hopping as well as the $\lambda_{mh}$ perturbation term. The second term corresponds to the effect of the HBN substrate which is a staggered mass on each layer with $\sigma_{z}$ denoting the sublattice degree of freedom. The last term corresponds to the interlayer hopping. We take a simple interlayer hopping model that is a function of the in-plane distance between two atoms in different layers and falls off exponentially with increasing distance. This gives the following form for the interlayer hopping, $t^{\perp}(r)=t_{v}\exp(-r/\eta)$, where $r$ is the in-plane distance between the two atoms, $\eta$ controls the range of the hopping and $t_v$ is the hopping amplitude. Lattice relaxations reduce the A-A hopping between the layers in contrast to A-B hopping~\cite{lattice_relaxation,PhysRevB.99.195419,carr2019exact,cantele2020structural} which we include in the model by the parameter $\kappa$ which defines the ratio between the two. 
\begin{figure}[t!]
\captionsetup[subfigure]{labelformat=empty}
\includegraphics[width=0.48\textwidth]{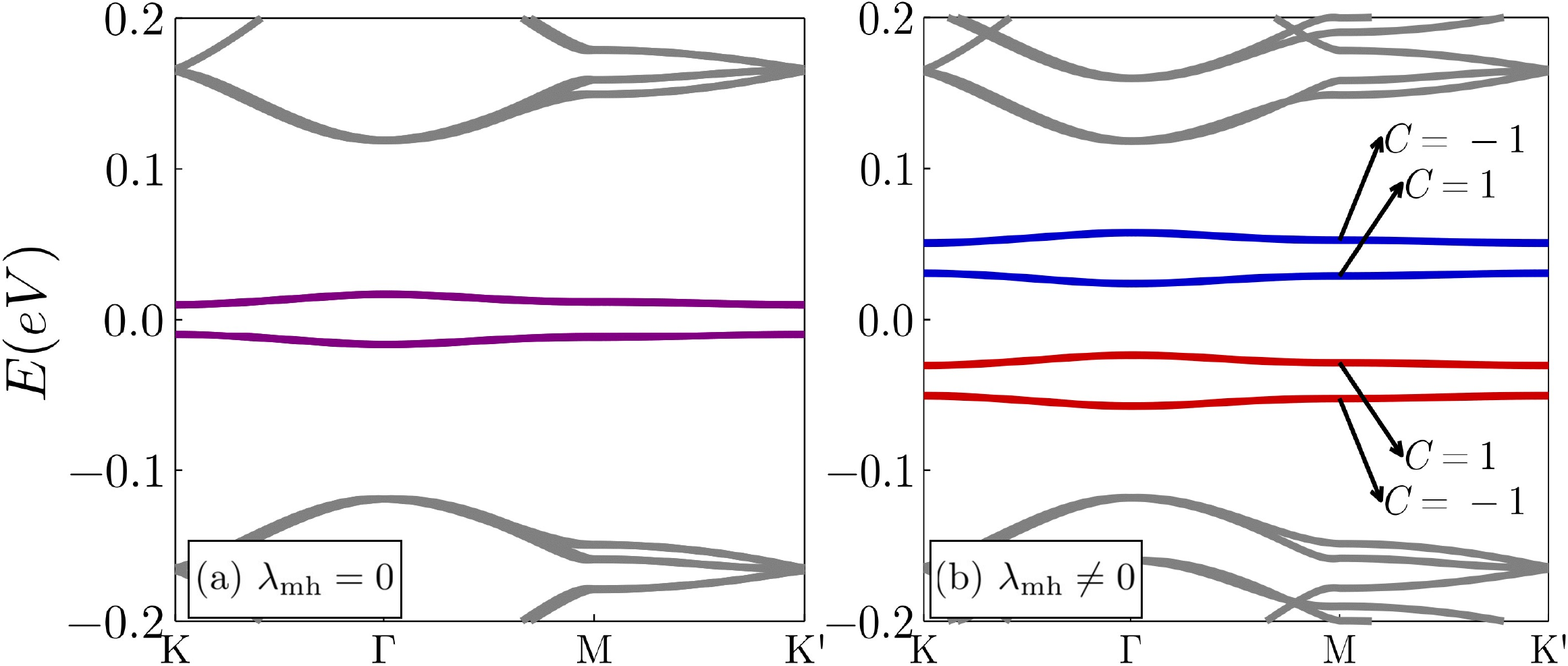}
	\caption{\label{commen_bands} The band structure of the commensurately twisted lattice model $H_{\rm TBG}$, Eq.~(\ref{eq:h}) in the moir\'e BZ computed with periodic boundary conditions (without edges). (a) Without the perturbation term, $\lambda_{\rm mh}=0$, ${\cal T}$ is preserved and a Chern insulator is not possible. The two isolated flat bands in the middle are doubly degenerate, which can be understood from the continuum model as arising due to the valley degeneracy. The remote bands in gray do not play an important role in our work. (b) Turning on $\lambda_{\rm mh}$ breaks ${\cal T}$ and allows the moir\'e system to have bands that have well defined Chern numbers, the arrows point to the value of the Chern number of the bands. Chern numbers are computed by the standard lattice method~\cite{doi:10.1143/JPSJ.74.1674}. This spinless Moire system would be a Chern insulator at 1/4 and 3/4 filling of the flat band system. We used $\theta\approx1.08^\circ$, $t=2.7 eV$, $t_v=0.686 eV$, $\kappa=0.0$, $\eta=0.3a$, $m_t=m_b=10meV$ and $\lambda_{\rm mh}=8meV$.}
\end{figure}

In Fig. (\ref{commen_bands}) we show the band structure of the model first with $\lambda_{\rm mh}=0$ and then with $\lambda_{\rm mh}\neq 0$. For $\lambda_{\rm mh}=0$, there are two flat bands that are each doubly degenerate. The degeenracy can be understand in the continuum model, as arising from the two valleys.   Since ${\cal T}$ is preserved, no Chern insulator is possible. For $\lambda_{\rm mh}\neq0$ we observe that the flat bands split into four isolated bands (one pair moves up and the other pair moves down in energy). As we would expect from the continuum model picture, since the $\lambda_{\rm mh}$ perturbation raises the energy of one valley's Dirac touching with respect to the other in monolayer  graphene, the two flat bands in one valley move up, while those in the other move down. The Chern numbers of the pair that move up are $\pm 1$, while those of the pair that move down have $\mp 1$, consistent with the fact that they arose from two valleys connected by time reversal symmetry when $\lambda_{\rm mh}=0$. Evidently, if we fill the flat band manifold with only enough electrons to fill one band (or three bands), we obtain a Chern insulator. This demonstrates that the $\lambda_{\rm mh}$ perturbation can turn the moir\'e system into a quantum anomalous Hall insulator.  We note here for completeness that for a single sheet of graphene, $H_{\rm mh}$ supplemented with a sublattice mass term breaks ${\cal T}$ but nonetheless gives rise only to a trivial insulator;  the Chern insulator that we find here for the twisted case essentially requires the subtle moir\'e band reconstructions. 

\begin{figure}[t!]
\captionsetup[subfigure]{labelformat=empty}
\includegraphics[width=0.5\textwidth]{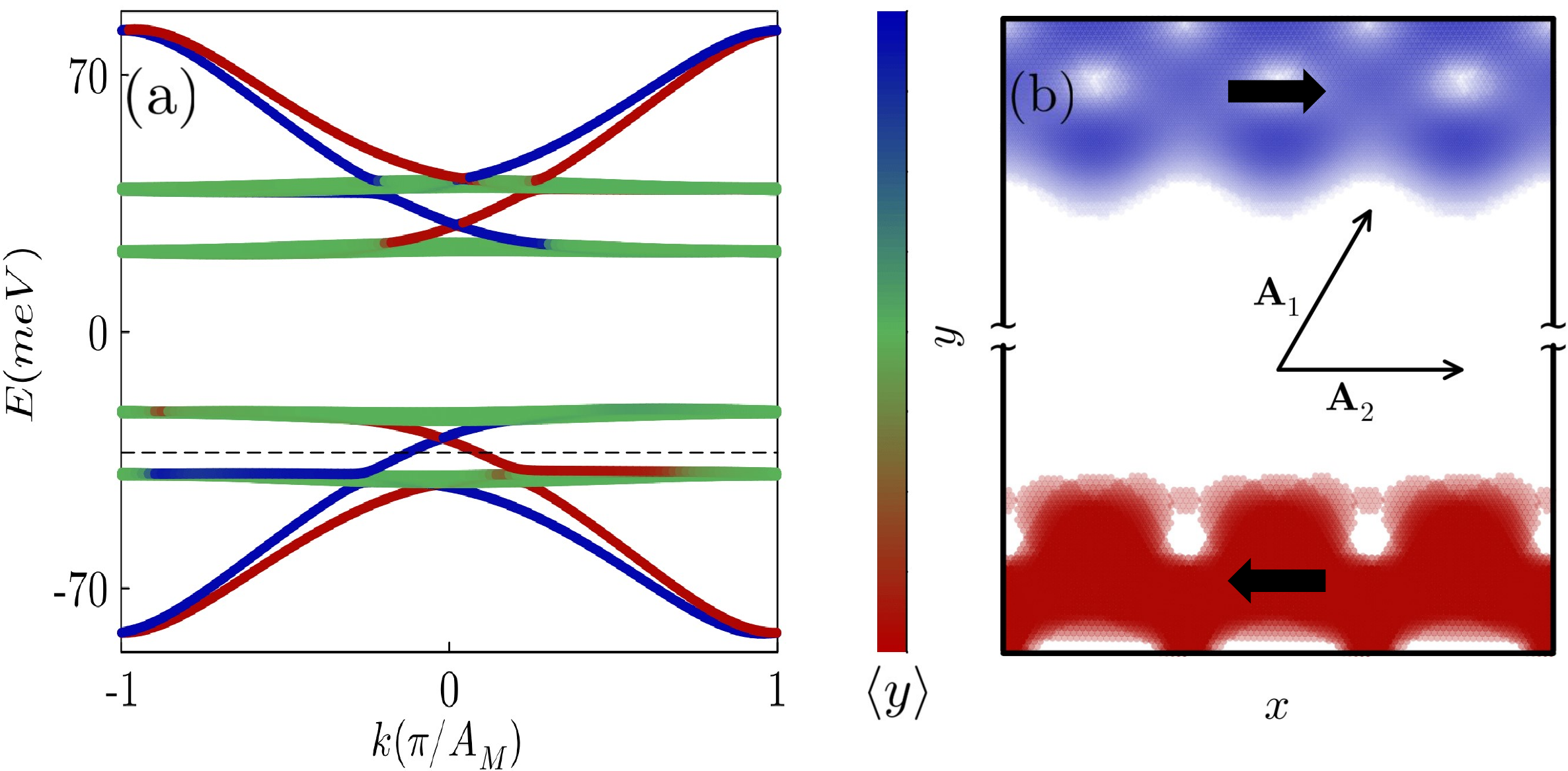}
	\caption{\label{rib_bands}  Band structure and wavefunctions of $H_{\rm TBG}$ in a ribbon geometry (with edges) with the same parameters as in Fig.~\ref{commen_bands} and with a transverse width of 20 moir\'e unit cells and infinite longitudinal direction. (a) Shows the bands in a window of energy that includes the four flat bands of Fig.~\ref{commen_bands}(b). The color of the eigenstates represents the expectation value of the transverse co-ordinate of the ribbon, in accordance with the shown color bar. The bulk bands appear green since their average transverse co-ordinate is in the middle. Likewise edge states localized on the upper (lower) edge appear blue (red). We observe that chiral mid-gap states show up between the first and second and between the third and fourth flat bands, corresponding to the edge states of a Chern insulator at 1/4 and 3/4 filling. In addition, counter-propagating edge states not associated with the QAHE appear below the first band and above the fourth band. A possible fermi level for a Chern insulator at bulk filling of 1/4 is shown as a dashed line. (b) Shows the detail of $|\Psi|^2$ for the two edge states' wave functions at the Fermi level marked in (a). The system diagonalized had a transverse width of 20 moir\'e unit cells, but for clarity most of the bulk has been omitted as represented by the broken $y$-axis. For visual clarity we have shown three repeats of the unit cell along the longitudinal direction. Black arrows indicate the chirality of the edge modes inferred from their group velocities in (a).} 
\end{figure}

{\em Edge States:} Having established that the $\lambda_{\rm mh}$ perturbation creates a QAH state from the bulk $k$-space topology, we now study the model in a ribbon geometry to access the edge states. We construct a ribbon by taking the system to be infinite along the ``longitudinal" $\mathbf{A}_2$ direction (with the momentum parallel to $\mathbf{A}_2$ being a good quantum number) and finite along the ``transverse" $\mathbf{A}_1$ direction which defines the width of the ribbon. Due to the large number of atoms in the unit cell we use the Lanczos algorithm to obtain the bands near charge neutrality where the Chern bands and hence the edge states reside. Fig. (\ref{rib_bands})(a) shows the band structure of the Hamiltonian in Eq.~(\ref{eq:h}) of a ribbon with a width of 20 moir\'e unit cells. We see the projection of the four bulk bands near charge neutrality but most importantly we see two pairs of gap crossing modes that are localized on the upper and lower edges of the ribbon that correspond to the edge states of the Chern insulator. The location of the support of the wavefunctions in the transverse direction is determined by computing the position expectation value of the wave function along the width of the ribbon~\cite{PhysRevB.103.155410}. Note that there are also edge states that lie at energies above and below the four flat bands which are reminiscent of the ``moir\'e edge states" that exist before adding the $H_{\rm mh}$ perturbation~\cite{PhysRevB.87.075433,PhysRevB.89.205405,PhysRevB.91.035441,PhysRevB.99.155415,PhysRevB.97.205128,PhysRevB.103.155410,Ma2020}. These states  are counter propagating along each edge which makes them susceptible to back scattering in contrast with the robust edge states of the Chern insulator discussed here. Furthermore, they do not cross the band gap between the flat bands and the remote bands. 

What is the transverse localization length of the edge states? Given the microscopic form of the mean field perturbation in Eq.~(\ref{eq:hmh}) (illustrated in Fig.~\ref{fig:hmh}(a)) one may naively assume that the edge states  are localized on the scale of a graphene lattice spacing. On the other hand, in the continuum model the moir\'e lattice spacing $A_M$ is the only length scale that appears, and hence we might expect that the edge state wave function to be localized on this larger scale. We plot the edge state wave function squared amplitude in Fig.~\ref{rib_bands}(b). We see that the edge mode wavefunctions extend up to approximately one moir\'e unit cell length $A_M$, in agreement with the expectation from the continuum model. The fundamental graphene lattice constant $a$ is about 50 times smaller than $A_M$. Furthermore, we have checked that the edge state localization length depends linearly on $A_M$ demonstrating that it is tied to this larger length scale (See supplemental material).  

\begin{figure}[t!]
\includegraphics[width=0.49\textwidth]{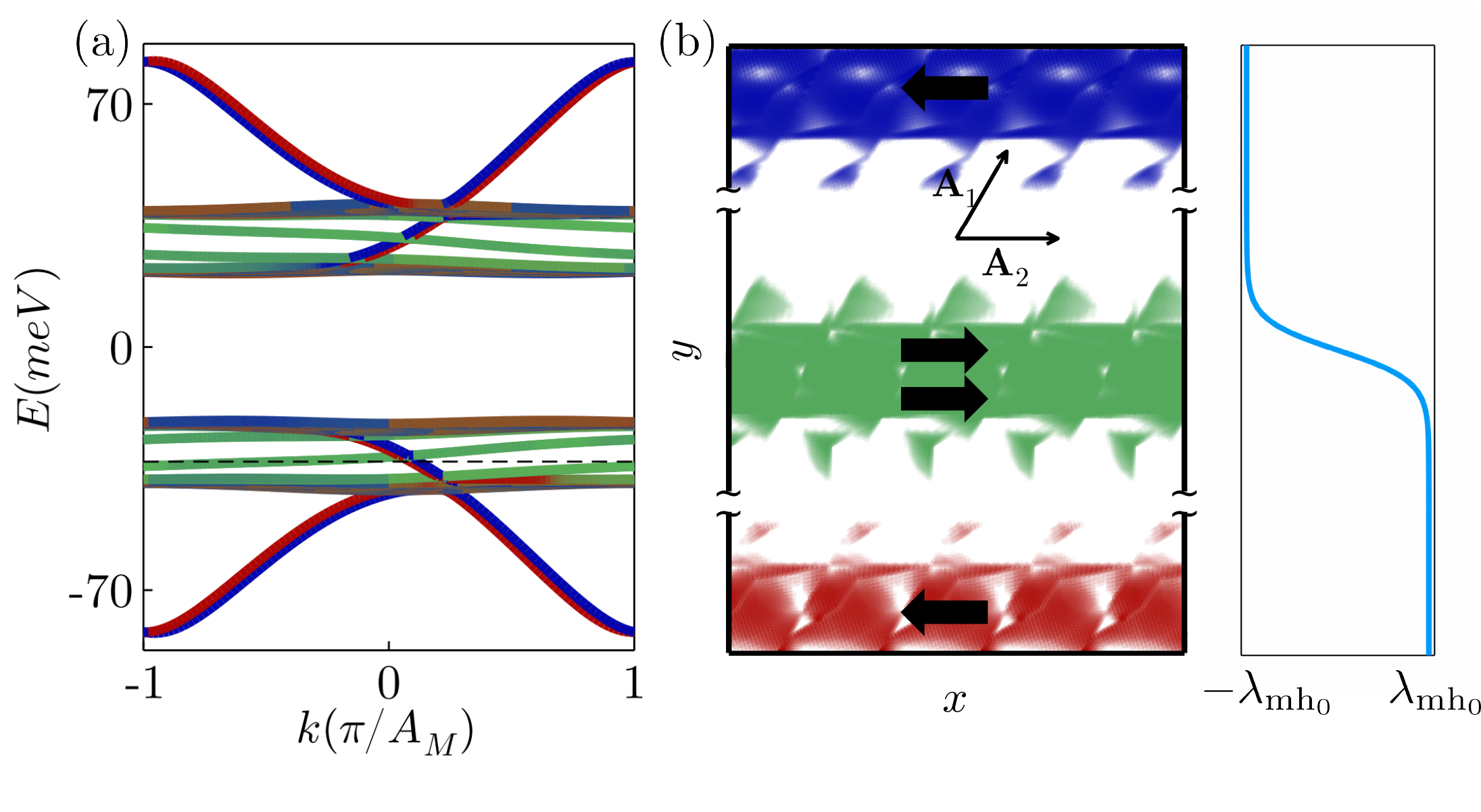}
		
	\caption{\label{domain} Electronic structure of a QAHE domain wall in the ribbon geometry. (a) Shows the band structure in the presence of a $C=\pm 1$ domain wall (see text), colored with the same colorbar as in Fig.~\ref{rib_bands}. At the Fermi level (dashed line) for 1/4 bulk filling there are now four chiral edge states: two of these are quasi-degenerate and localized in the middle (and hence appear green) at the domain wall, and the other two are at the upper and lower edge of the ribbon. (b) Density for the localized mode wave functions at the 1/4 Fermi level, showing the chiralities inferred from the group velocities of the band structure. For clarity we show only one of the two modes trapped on the domain wall, since they have similar support. We have shown five repeats along the longitudinal direction. The profile of the domain wall used ($\xi=0.9 A_M$) is shown on the right.  All other parameetrs are same as in Fig.~\ref{commen_bands}.}
\end{figure}

{\em Domain Walls:}
The QAHE phase in moir\'e graphene has been argued to be an orbital ferromagnet, which breaks time reversal symmetry spontaneously~\cite{andrei2020graphene,liu2021orbital,PhysRevB.99.075127,PhysRevLett.124.166601,PhysRevLett.124.046403,PhysRevLett.124.187601,PhysRevB.103.075122,PhysRevB.104.115404,PhysRevLett.125.226401,PhysRevLett.124.046403,PhysRevB.103.035427,PhysRevResearch.1.033126,PhysRevResearch.3.013033,PhysRevB.102.035136,PhysRevB.100.085136,PhysRevB.102.045107,PhysRevB.102.035441}. This discrete symmetry breaking results in two ground states corresponding to clockwise or anti-clockwise chiral edge states ($C=\pm 1$) connected by time reversal symmetry, which can be accessed in our model by choosing the mean field term $\lambda_{mh}$ either positive or negative. Experiments on twisted bilayer graphene aligned with hBN have reported the co-existence of ferromagnetic domains in the system\cite{doi:10.1126/science.aay5533,doi:10.1126/science.aaw3780,Chen2020,Grover2022,doi:10.1126/science.abd3190}, with a possible mesoscale pattern of opposite domains separated by domain walls. The domain walls themselves are fascinating objects that provide an experimentally tunable realization of the zero modes trapped by defects in topological band systems, a fundamental problem in physics~\cite{jr1976,sshrmp}.  Our model allows us to explicitly compute the wavefunctions of the chiral modes localized at the domain walls. To set up a domain wall, we study a configuration where the order parameter $\lambda_{mh}$ varies smoothly across the width of the ribbon between two regions of opposite Chern number. We use $\lambda_{mh}=\lambda_{\rm mh_{0}}\tanh{(y/\xi)}$, with $\xi$ a characteristic length scale of the domain wall. In Fig.~\ref{domain}(a) we show the resulting band structure. Focusing at a Fermi level corresponding to 1/4 filling (shown as a dashed line), we find four localized states: two states at the center of the sample (trapped by the domain wall), and one each on the upper and lower edges. The two parallel chiral states in the domain wall are consistent with the behavior of a domain wall separating regions with $C=\pm 1$ Chern numbers. The corresponding wave function for one of the domain wall states is shown in Fig.~\ref{domain}(b) exhibiting a state that is localized in the middle of the ribbon with a localization length controlled by $A_M$, the moir\'e length scale. Also shown in this figure is the upper and lower chiral edge state. 

{\em Energetics:} We have demonstrated that the $\lambda_{\rm mh}$ perturbation of Eq.~(\ref{eq:hmh}) does indeed create a mean field QAH state in moir\'e graphene, and the chiral edge modes appropriate for both a hard edge and a domain wall. Our mean field ansatz is also conceptually attractive, because it has only two ``moving parts": The sublattice masses to gap the flat bands and give them nontrivial Chern numbers in each valley, and $\lambda_{\rm mh}$ to raise one valley with respect to the other, which we have made simply arguments has to result in a Chern insulator.  However, it is interesting to step back and ask whether this mean field ansatz is unique or favored in some sense.  From a symmetry point of view, the only necessary (but not sufficient) requirement to get a Chern insulator is that the perturbation break ${\cal T}$. Clearly many imaginary hopping patterns, such as Haldane's perturbation, or those involving further neighbor hopping and modulation of the hopping  with the moir\'e periodicity,  can achieve this. However, unlike our simple ansatz, it is a priori unclear which of these will result in a Chern insulator. A full Hartree-Fock calculation on the lattice, including all the bands, will by definition lead to the best mean field Hamiltonian. However, this is computationally prohibitive. To gain further physics insight into the issue in a limited parameter space, we have constructed a mean field model which has the sublattice masses, the modified Haldane term with a coefficient $\lambda_{\rm mh}$, and also a Haldane term with a coefficient $\lambda_{\rm h}$. Note that the Haldane term by itself also leads to a Chern insulator for TBG at quarter filling. We variationally compute the total energy, including the Coulomb interaction screened by gates~\cite{andrei2020graphene,liu2021orbital,PhysRevB.99.075127,PhysRevLett.124.166601,PhysRevLett.124.046403,PhysRevLett.124.187601,PhysRevB.103.075122,PhysRevB.104.115404,PhysRevLett.125.226401,PhysRevLett.124.046403}, as a function of $\lambda_{\rm mh},\lambda_{\rm h}$ for a fixed set of sublattice masses. As shown in the SM, we find that the minimum occurs when $\lambda_{\rm h}=0, \lambda_{\rm mh}\neq0$. Thus, at least in this limited set of variational parameters, the modified Haldane term is preferred.

{\em Conclusion:}
In summary we have introduced a lattice mean field model for the quantum anomalous  Hall effect in twisted bilayer graphene. The mechanism that gives rise to the Chern insulator is physically appealing: The sublattice masses on the two layers break ${\cal C}_2$, gap out the flat bands in each valley and give them nontrivial Chern numbers, while the modified Haldane term breaks ${\cal T}$ and raises the states in one valley with respect to the other. Our mean field model's conceptual simplicity, its computational tractability, and its good energetics (albeit in a limited parameter space) make it attractive for study. By solving our lattice model in a ribbon geometry, we find that the chiral edge modes of the QAH state have a transverse localization length that scales with the moir\'e lattice spacing. We have also used our model to study the domain walls occurring between regions where the order parameter changes sign.

A number of interesting open questions involving spatially inhomogeneous systems  can be studied using our lattice approach. 
The control of the electronic structure of the chiral modes by the shape of the domain wall, a study of  networks of domain walls and the effect of impurities are some examples. 
%A second example is effect of superconducting contacts, which would be expected to proximitize TBG, and perhaps lead to reconstructions in the chiral edge modes. 
We hope to study these and other questions in the near future. 

{\em Note Added:} While our manuscript was being finalized for submission, an experimental work using STM on domain walls in twisted monolayer-bilayer graphene has appeared, in which the QAHE edge states have been imaged for the first time~\cite{aqhe_stm}. Although closely related to the system under study here, it is different in detail. The main idea of our theory can be adapted straightforwardly to this work.

{\em Acknowledgments:}
This work was supported in part by NSF DMR-2026947 (AK, RKK). GM is grateful to the US-Israel BSF for partial support under grant no. 2016130. GM and RKK acknowledge the Aspen Center for Physics, NSF PHY-1607611 (GM, RKK) for its hospitality. GM is also grateful to the International Center for Theoretical Sciences, Bangalore for its hospitality while this work was being completed. The authors are grateful to the University of Kentucky Center for Computational Sciences and Information Technology Services Research Computing for their support and use of the Lipscomb Compute Cluster and associated research computing resources.

\bibliography{ref}
\end{document}

\title{Supplementary Material for ``Lattice Model For The  Quantum Anomalous Hall Effect in Moir\'e Graphene''}
\author{Ahmed Khalifa}
\affiliation{Department of Physics \& Astronomy, University of Kentucky, Lexington, KY 40506, USA}
\author{Ganpathy Murthy}
\affiliation{Department of Physics \& Astronomy, University of Kentucky, Lexington, KY 40506, USA}
\author{Ribhu K. Kaul}
%\affiliation{Department of Physics \& Astronomy, University of Kentucky, Lexington, KY 40506, USA}
\affiliation{Department of Physics, The Pennsylvania State
University, University Park, PA 16802, USA}
\maketitle
In this set of supplemental materials we give details about the lattice setup we have used to build our mean field model and we also present calculations
substantiating statements made in the main text. In section \ref{SM:sec1} we review commensurate twisted bilayer graphene which has exact periodicity for a discrete set of twist angles. Such model allows us to use Bloch theorem to get the band structure and the wavefunctions of the edge and domain wall states in the QAH phase. In section \ref{SM:sec2} we present calculations of the edge states localization length. We support our claim that the edge state localize on the large moir\'e scale. As we have computed the localization length at different twist angles, there is a dircet linear relationship between the moir\'e length scale and the edge state localization length. Finally in section \ref{SM:sec3} we present calculations of the mean field state total energy which includes the kinetic and the screened Coulomb interaction between the electrons. Our numerical data supports the claim that the state obtained by $\lambda_{\rm mh}$ is preferred energetically to the state obtained by $\lambda_{\rm h}$.  
\section{Twisted bilayer graphene Lattice Setup}
\label{SM:sec1}
In this section we give details about the lattice construction that we use in building our model. In order to have true periodicity in real space we study commensurate twisted bilayer graphene. Although a general twist angle may result into a system that is not periodic, there exists an infinite set of angles where the twsited bilayer graphene system is exactly periodic. This set of discrete twist angles is defined by co-prime positive integers $(m,n)$ that is given by the formula, 
\begin{equation}\label{eq2}
    \cos{\theta}=\frac{m^2+n^2+4mn}{2(m^2+n^2+mn)}
\end{equation}
The system now forms a superlattice that gets bigger as the twist angle gets smaller with superlattice translation vectors that are defined in terms of the monolayer lattice vectors as, 
\begin{equation}
\mathbf{A}_1=n\mathbf{a}^t_1+m\mathbf{a}^t_2, \mathbf{A}_2=-m\mathbf{a}^t_1+(m+n)\mathbf{a}^t_2,
\end{equation}
where $\mathbf{a}^t_{1,2}=R(\theta/2)\mathbf{a}_{1,2}$ are the translation vectors of the top graphene layer with $R(\theta/2)$ being the 2D matrix of rotation and $\mathbf{a}_{1,2}$ are the unrotated graphene lattice translation vectors. The length of the moir\'e lattice translation vectors is expressed in terms of the rotation angle as,
\begin{equation}
    A_M=\abs{\mathbf{A}_1}=\abs{\mathbf{A}_2}=\frac{(m-n)a}{2\sin{\theta/2}},
\end{equation}
where $a$ is the length of the graphene lattice constant with the assumption that $m>n$ and we consider cases with $m-n=1$ where the moir\'e and the superlattice periodicity coincide~\cite{PhysRevB.81.161405,PhysRevB.81.165105}.    

\section{Localization length scale of the QAH edge states}
\label{SM:sec2}
In this section we investigate the transverse localization of the topological edge states in the QAH phase of twisted bilayer graphene obtained via our mean field model. In order to answer this question, we introduce a mechanism to define what is meant by the transverse localization length and then examine how the localization length depends on the twist angle $\theta$, and consequently the moir\'e length scale $A_M$. 

\begin{figure}[h!]
	\subfloat[\label{locvsE}]{	
  \includegraphics[width=0.45\textwidth]{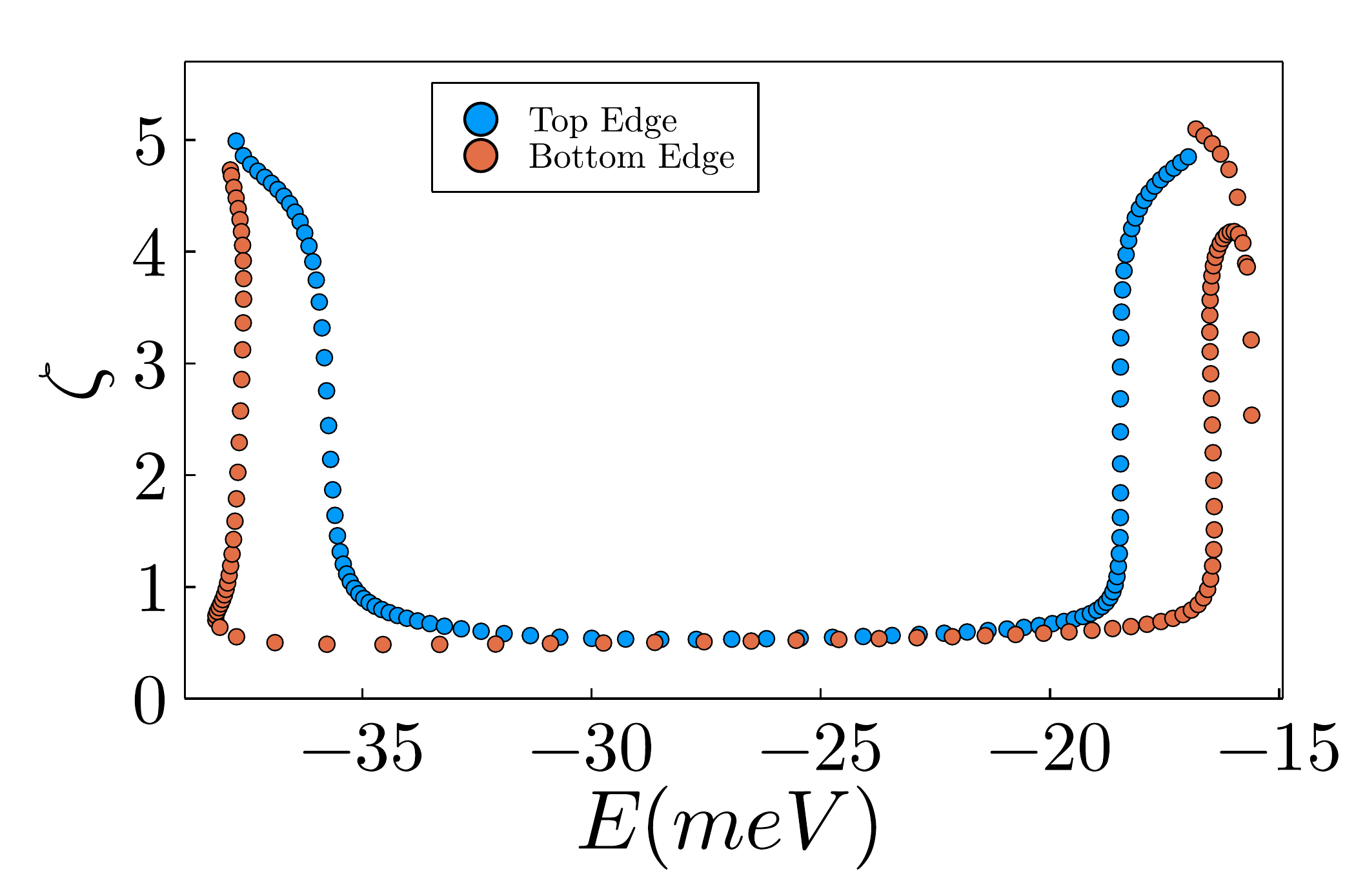}}
  \hspace{0.05em}
	\subfloat[\label{locvsmoire}]{%
		\includegraphics[width=0.45\textwidth]{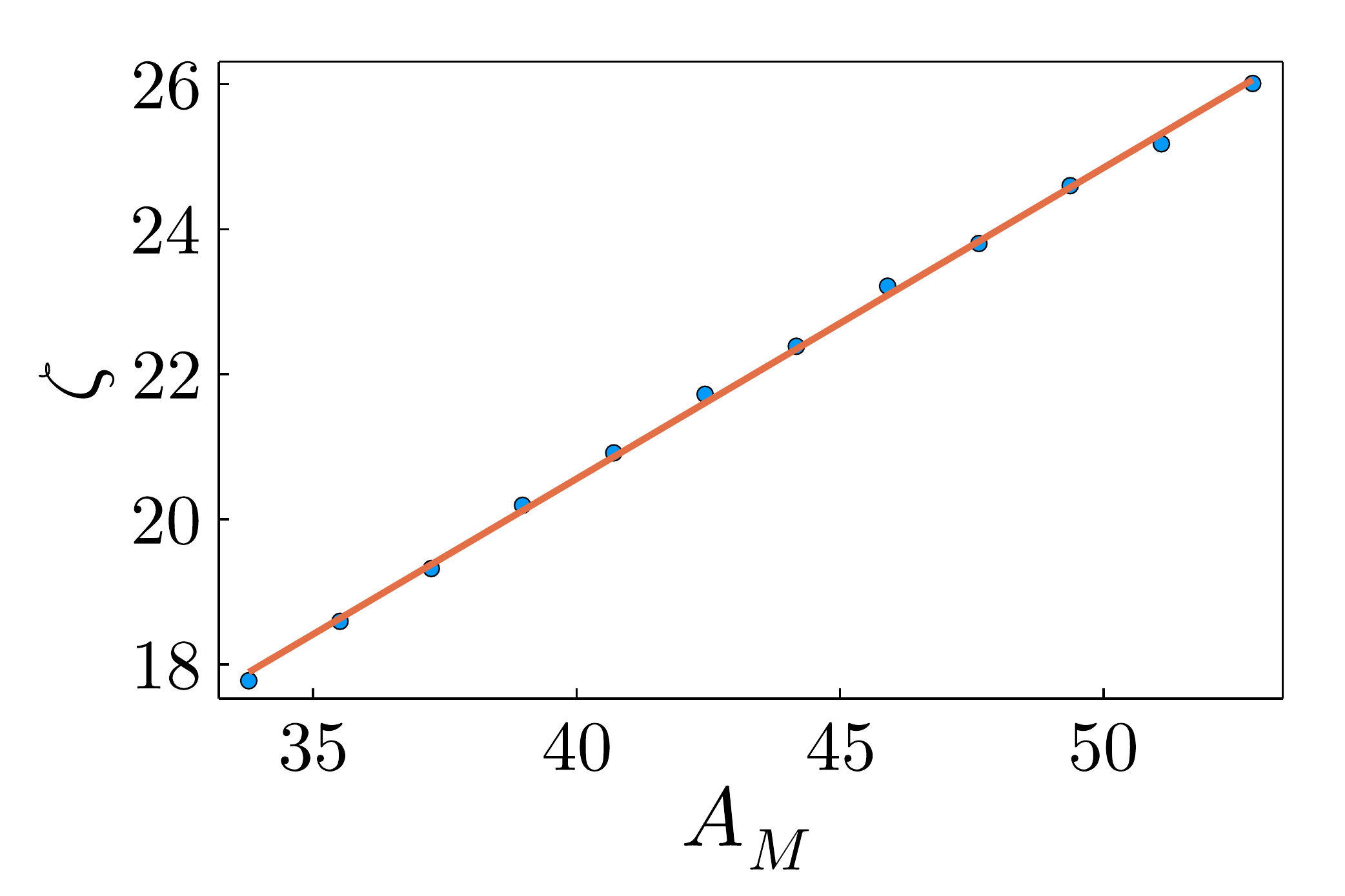}}
	\caption{\label{loc} The Dependence of the edge states localization length on the twist angle. (a) Shows the dependence of the edge states belonging to the top/bottom edges on their energies. The ``U'' shape is a result of the edge states getting spread out as they mix with the bulk states and being localized away from the bulk states. (b) Shows the edge states localization length $\zeta$ vs. the length of the moir\'e lattice vector $A_M$ (Both expressed in units of the monolayer graphene lattice constant $a$). The data fits a straight line as a direct indication of the edge states being localized on the moir\'e scale.}
\end{figure}

An edge band lies in the gap between two bulk bands with nonzero Chern numbers and its energy disperses in momentum connecting the two bulk bands. When the edge band get ``absorbed'' by the bulk bands, the edge states localization length diverges and the state becomes spread out in the bulk. Given the wave function of an edge state (obtained by the diagonalization of the system on a ribbon as illustrated in the main text), its localization length is defined by computing the position expectation value along the transverse direction for the states localized at the bottom of the ribbon and for the states localized at the top edge it is given by the difference between the ribbon width and the position expectation value. Plotting the localization length states belonging to an edge band against their energies makes a ``U'' shape ( See Fig. (\ref{locvsE}))corresponding to them being spread out as their energy approaches the bulk bands and getting localized as edge states away from the bulk bands. We associate a localization length with an edge band with the statistical average of its edge states. Our numerical analysis tells us two things: (i) The localization length of the edge states is of the order of one moir\'e lattice vector length. (ii) There is a linear relationship between the localization length and the moir\'e length scale (See Fig. (\ref{locvsmoire})). This confirms the expectation that edge states of the QAH phase are localized on the moir\'e scale which would be interesting to show in an STM experiment that can probe the edge state wave function in real space. 
\section{Energy of the mean field state with electron-electron interactions}
\label{SM:sec3}
\begin{figure}[h!]
	%\subfloat[\label{locvsE}]{	
  \includegraphics[width=0.45\textwidth]{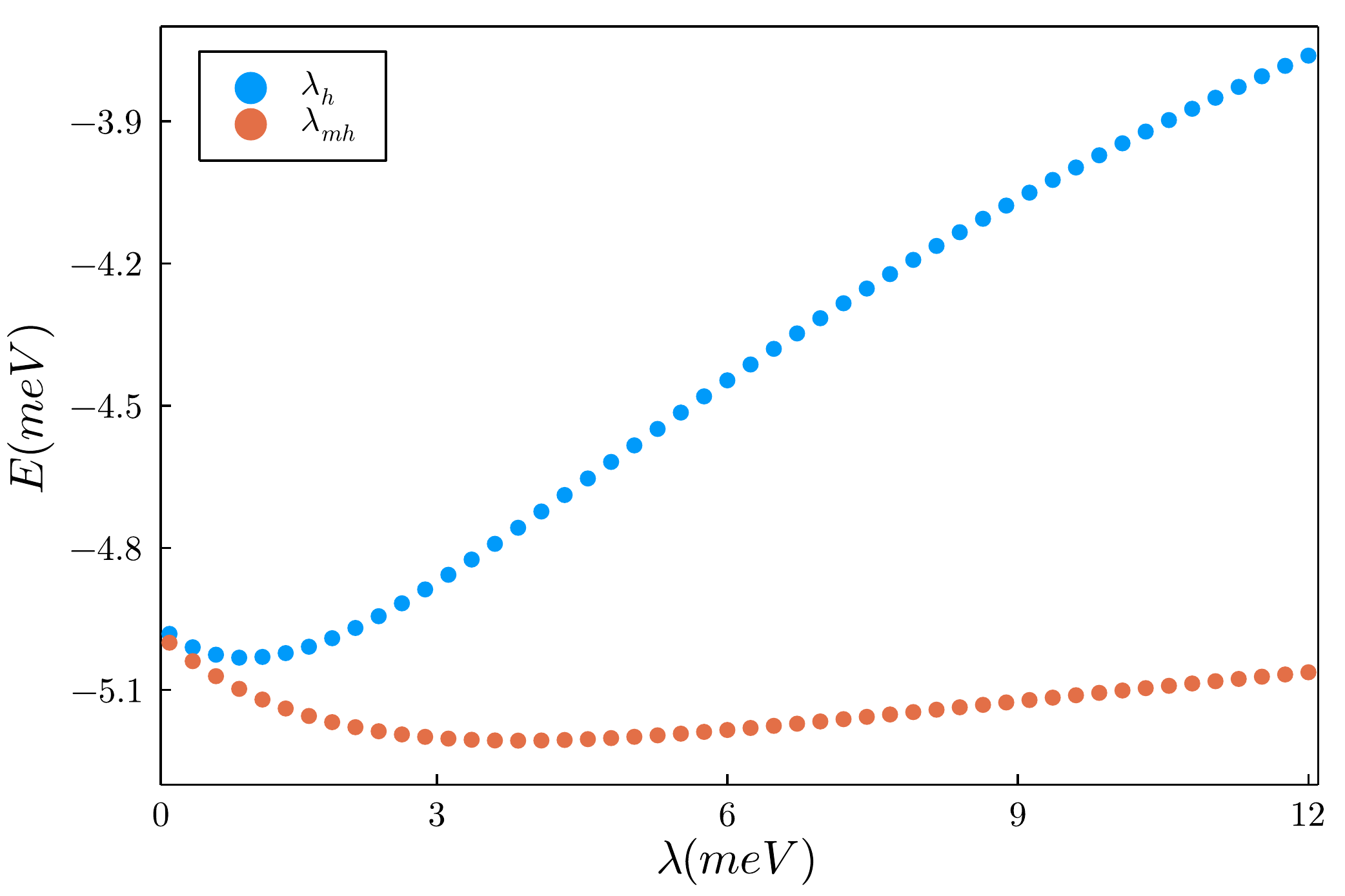}
  %\hspace{0.05em}
	%\subfloat[\label{locvsmoire}]{%
		%\includegraphics[width=0.45\textwidth]{localization_vs_moire.pdf}}
	\caption{\label{en} The total energy of the QAH mean field state obtained using either $\lambda_{\rm mh}$ or $\lambda_{\rm h}$ with fixed sublattice masses. The total energy of the state obtained by using only $\lambda_{\rm mh}$ is lower than the one obtained using only $\lambda_{\rm h}$ which means that the $\lambda_{\rm mh}$ state is preferred (at least in this limited set
of variational parameters). We have have used the following parameters $\theta\approx1.08^\circ$, $t=2.7 eV$, $t_v=0.686 eV$, $\kappa=0.0$, $\eta=0.3a$, $m_t=m_b=10meV$, $d=2~A_M$ and $\epsilon=5\epsilon_0$}
\end{figure}

In this section we show the details of calculating the energy of the mean field state which is made out of two parts, kinetic and electrons repulsion energy. First, we note that our mean field ansatz is parameterized by two parameters $\lambda_{\rm mh}$ and $\lambda_{\rm h}$. We obtain a QAH state by filling the bands up to the first moir\'e flat band which has a nonzero Chern number found by diagonalizing $H_{\rm TBG}$ defined in equation (2) in the main text. We first show how to compute the interaction energy. We start with writing the electron's field operator expanded in a basis spanned by a set of bands, 
\begin{equation}
    \label{eq:fieldop}
    \Psi(\mathbf{r})=\sum_{n,\mathbf{k}}\frac{1}{\sqrt{S}}e^{i\mathbf{k}.\mathbf{r}}u_{n\mathbf{k}}c_{n\mathbf{k}}, 
\end{equation}
where $n$ denotes the band index and $c_{n\mathbf{k}}$ denotes the electron's annihilation operator at state $\mathbf{k}$ and band $n$, and $S$ denotes the area of the sample. $u_{n\mathbf{k}}$ are the eigenvectors obtained by diagonalizing the single particle Hamiltonian $H_{\rm TBG}$.  
Through equation (\ref{eq:fieldop}) we can define the electron density operator as,
\begin{equation}
    \label{eq:densityop}
    \rho(\mathbf{r})=\Psi^{\dagger}(\mathbf{r})\Psi(\mathbf{r})= \sum_{m,n,\mathbf{k},\mathbf{q}}\frac{1}{S}e^{i(\mathbf{q}-\mathbf{k}).\mathbf{r}}u^{\dagger}_{n\mathbf{k}}u_{m\mathbf{q}}c^{\dagger}_{n\mathbf{k}}c_{m\mathbf{q}}.
\end{equation}
The electron-electron interaction is given by,
\begin{equation}
    \label{eq:intr}
    V=\int d^{2}\mathbf{r}_{1}d^{2}\mathbf{r}_{2} v(\mathbf{r}_1-\mathbf{r}_2)\rho(\mathbf{r}_1)\rho(\mathbf{r}_2).
\end{equation}
 We can write the interaction in momentum space via a Fourier transform,
 \begin{equation}
     \label{eq:intk}
     V=\sum_{\mathbf{k}}v(\mathbf{k}):\rho(\mathbf{k})\rho(-\mathbf{k}):,
 \end{equation}
 where $v(\mathbf{k})$ is the Fourier transform of $v(\mathbf{r}_1-\mathbf{r}_2)$. For a Coulomb interaction screened by the gates $v(\mathbf{k})$ takes the form $v(\mathbf{k})=\frac{e^2\tanh{kd}}{2\epsilon k}$ where $d$ denotes the screening length via the metal gates which is about $40~\text{nm}$ which is about a few moir\'e lattice vector lengths at the twist angle $\theta=1.08^\circ$. Going back to equation (\ref{eq:intk}) we can write $\rho(\mathbf{k})$ as,
 \begin{equation}
     \label{eq:rhok}
     \rho(\mathbf{k})=\sum_{\mathbf{r}}e^{-i\mathbf{k}.\mathbf{r}}\rho(\mathbf{r})=\sum_{\mathbf{r}}e^{-i\mathbf{k}.\mathbf{r}}\sum_{m,n,\mathbf{k},\mathbf{q}}e^{i(\mathbf{q}-\mathbf{k}).\mathbf{r}}u^{\dagger}_{n\mathbf{k}}u_{m\mathbf{q}}c^{\dagger}_{n\mathbf{k}}c_{m\mathbf{q}}.
 \end{equation}
 Doing the sum over $\mathbf{r}$ gives the following,
 \begin{equation}
     \rho(\mathbf{k})=\sum_{m,n,\mathbf{q},\mathbf{G}}u^{\dagger}_{n,\mathbf{q}}u_{m,\mathbf{q}+\mathbf{k}+\mathbf{G}}c^{\dagger}_{n,\mathbf{q}}c_{m,\mathbf{q}+\mathbf{k}},
 \end{equation}
 where $\mathbf{G}$ denotes a moir\'e reciprocal lattice vector. While for now we have been keeping the sum over the bands, we want to project onto the flat bands which takes care of the sum over the bands indices. This gives the following for the electron electron interaction term,
 \begin{equation}
     \label{eq:projint}
     V=\sum_{\mathbf{k},\mathbf{q}_1,\mathbf{q}_2}v(\mathbf{k})c^{\dagger}_{\mathbf{q}_1}c^{\dagger}_{\mathbf{q}_2}c_{\mathbf{q}_2-\mathbf{k}}c_{\mathbf{q}_1+\mathbf{k}}u^{\dagger}_{\mathbf{q}_1}u_{\mathbf{q}_1+\mathbf{k}}u^{\dagger}_{\mathbf{q}_2}u_{\mathbf{q}_2-\mathbf{k}}
 \end{equation}
 Given this form of the interaction and the QAH state obtained by our mean field Hamiltonian, we can compute its energy by taking its expectation value which has both Hartree and Fock terms. The Hartree term cancels with the back ground charge and what remain is the Fock term,
 \begin{equation}
     \label{eq:fock}
     V^F=-\frac{1}{2S}\sum_{\mathbf{k},\mathbf{q}}v(\mathbf{k})|\langle u(\mathbf{q})| u(\mathbf{q}-\mathbf{k})\rangle|^2.
 \end{equation}
 Note the sum in equation (\ref{eq:fock}) is over all momenta and is not restricted to the Brillouin zone. We can also express it as,
 \begin{equation}
     \label{eq:fock2}
     V^F=-\frac{1}{2S}\sum_{\mathbf{k},\mathbf{q}\in\text{BZ}}\sum_{\mathbf{G}}v(\mathbf{k}+\mathbf{G})|\langle u(\mathbf{q})| u(\mathbf{q}-\mathbf{k})\rangle|^2.
 \end{equation}
 The kinetic energy term is easily computed by getting the single particle Hamiltonian expectation value in the QAH state. Our numerical calculation for the energy of the QAH state shows that the state obtained by using the $\lambda_{\rm mh}$ perturbation has lower energy than the one obtained by $\lambda_{\rm h}$ (See Fig. (\ref{en})).
\bibliography{sup_ref}